\documentclass[aps,twocolumn,prl,nofootinbib]{revtex4}
\usepackage{graphicx}
\usepackage{color}

\newcommand{ \slashchar }[1]{\setbox0=\hbox{$#1$}   
   \dimen0=\wd0                                     
   \setbox1=\hbox{/} \dimen1=\wd1                   
   \ifdim\dimen0>\dimen1                            
      \rlap{\hbox to \dimen0{\hfil/\hfil}}          
      #1                                            
   \else                                            
      \rlap{\hbox to \dimen1{\hfil$#1$\hfil}}       
      /                                             
   \fi}                                             %

\def\lsim{\mathrel{\raise.3ex\hbox{$<$\kern-.75em\lower1ex\hbox{$\sim$}}}}
\def\gsim{\mathrel{\raise.3ex\hbox{$>$\kern-.75em\lower1ex\hbox{$\sim$}}}}

\begin{document}
\title{Emergent Electroweak Gravity}

\author{Bob McElrath}
\email[]{bob.mcelrath@cern.ch}
\affiliation{CERN theory group, Geneva 23, CH 1211, Switzerland}

\vskip 0.2in 
\begin{abstract}
    We show that any massive cosmological relic particle with small
    self-interactions is a super-fluid today, due to the broadening of
    its wave packet, and lack of any elastic scattering.  The WIMP dark
    matter picture is only consistent its mass $M \gg M_{\rm Pl}$ in
    order to maintain classicality.  The dynamics of a super-fluid are
    given by the excitation spectrum of bound state quasi-particles,
    rather than the center of mass motion of constituent particles.  If
    this relic is a fermion with a repulsive interaction mediated by a
    heavy boson, such as neutrinos interacting via the $Z^0$, the
    condensate has the same quantum numbers as the vierbein of General
    Relativity.  Because there exists an enhanced global symmetry
    $SO(3,1)_{space}\times SO(3,1)_{spin}$ among the fermion's
    self-interactions broken only by it's kinetic term, the long
    wavelength fluctuation around this condensate is a Goldstone
    graviton.  A gravitational theory exists in the low energy limit of
    the Standard Model's Electroweak sector below the weak scale, with a
    strength that is parametrically similar to $G_N$.  
\end{abstract}
\noindent 

\def\os{\overline{\sigma}}
\maketitle

\setcounter{footnote}{0}

\section{Introduction}

In the early universe, relics including photons,
neutrinos and dark matter evolve out of thermal equilibrium
as their interaction strength becomes small at low temperature in a
process known as ``freeze-out''.  This calculation is essentially
classical, assuming particles are point-like and using the Boltzmann
equation~\cite{griest_cosmic_1987,srednicki_calculations_1988}.

After freeze-out the number density of particles is fixed, and the
temperature just evolves with Hubble expansion.  Their time evolution is
given only by the free particle kinetic term.  It is usually assumed
that the interaction strength is so weak that it can be neglected and
that particles remain localized point particles forever.  The free
particle Hamiltonian propagates particles and also broadens their wave
packets, described by their uncertainty $\Delta x$.  This is due to the
fact that the localization of particles causes them to not be an
eigenstate of the Hamiltonian if they are massive.

There are two limits of interest for the particle uncertainty $\Delta x$
relative to the number density $n$.  The classical gas limit is $\Delta
x \ll n^{-1/3}$.  Elastic scattering collisions and the Boltzmann
equation describe this system.  The opposite limit, $\Delta x \gg
n^{-1/3}$ is a quantum liquid.  Because particles have wave function
overlap with their neighbors, one must take into account collective
effects due to contact interactions.  If there exists an attractive
interaction in any partial wave, then the vacuum energy can be lowered
by forming bound state quasi-particles.  The system will undergo a phase
transition to a super-fluid described by quasi-particles.

If the system contains global symmetries that are broken when the system
becomes a super-fluid, then Goldstone bosons will emerge.  As these are
massless, their dynamics are extremely important.  

The idea of gravity emerging from spinors is not new and fairly obvious,
as one can construct a spin-2 particle as the direct product of
spinors~\cite{ohanian_gravitons_1969,kraus_photons_2002}. However no
workable theory has been yet constructed.  The first idea of this type
is due to Bjorken~\cite{bjorken_dynamical_1963}, who attempted to
formulate the photon and graviton as a composite state.  The most recent
attempt and the most successful is due to Hebecker and
Wetterich~\cite{hebecker_spinor_2003,wetterich_gravity_2003}.
Their theory can be regarded as a reformulation of gravity in terms of
spinors, but they give no dynamics for the spinors which would lead to
such a theory.  This line of research
was largely killed by the paper of Weinberg and
Witten~\cite{Weinberg:1980kq}, which showed that a spin-2 particle could
not couple to a covariant conserved current.  Two ways out of this
theorem are to quantize geometry (the approach of string theory), or to
abandon diffeomorphism invariance as an exact symmetry.  Sakharov
originally suggested that the graviton could be emergent, and in such
theories, diffeomorphism invariance can only be
approximate~\cite{Sakharov:1967pk}.

\section{Quantum Liquid Transition}

The quantum liquid regime for a system occurs when the position
uncertainty $\Delta x$ is larger than the inter-particle spacing
\begin{equation}
    \Delta x \gg n^{-1/3}.
    \label{eq:quantumliquid}
\end{equation}
In this limit the system is not classical, and the condition of
scattering theory that the impact parameter $b \gg \Delta x$ cannot be
satisfied (often known as the ``well-localized'' assumption).

Particles in the classical gas limit will eventually time-evolve into a
quantum liquid in the absence of interactions.  The expansion of a free
particle wave packet in time is
\begin{equation}
    \Delta x(t)^2 = \Delta x_0^2 + \Delta v^2 t^2.
    \label{eq:deltaxt}
\end{equation}
This can be intuitively understood because different momentum components
may move with different velocities.  The wave number at $p+\Delta p$ has
a velocity $(p+\Delta p)/E$ while the wave number at $p-\Delta p$ has a
smaller velocity $(p-\Delta p)/E$ and these two wave numbers will
separate in space as they propagate if $E>p$.  

The condition for the time-independent 
super-fluid transition can be derived by neglecting the second term of
Eq.~\ref{eq:deltaxt}.  In the non-relativistic limit one arrives at
\begin{equation}
    T < \frac{\lambda^2 n^{2/3}}{3 m k_B}.
    \label{eq:nrtranstemp}
\end{equation}

The cross-section does not enter into this calculation, and the
uncertainty $\Delta x_0$ is assumed to be proportional to the thermal
de Broglie wavelength, $\Delta x_0 = 1/\Delta p = \lambda/p =
\lambda/\sqrt{3 m k T}$, where $\lambda$ is an $\mathcal{O}(1)$
parameter reflecting how ``localized'' the state is.  This temperature
may be further suppressed by elastic collisions, which must occur
frequently enough to keep particles localized to their thermal de
Broglie wavelength, but not so often that they destroy the condensate.

In the relativistic case, we also use Eq.~\ref{eq:deltaxt}, however the
velocity uncertainty for relativistic states is
\begin{equation}
    \Delta v = \frac{\Delta p}{E} (1-v^2)
    \label{eq:deltav}
\end{equation}
where $v=p/E$.  This correctly reflects the relativistic limit, $v \to
c$; massless wave packets do not broaden as each wave number propagates
with the same velocity, $v=c$.

The relevant time scale for wave packet broadening is the mean time
between collisions $\tau = 1/\sigma n v$ in terms of the cross section
$\sigma$ since the uncertainty of a
wave packet $\Delta x_0$ is set by the 3-momentum of an elastic
scattering collision.  The condition for a quantum liquid is then 
\begin{equation}
    \frac{1}{p^2} + \frac{(1-v^2)^2}{\sigma^2 n^2} > \frac{1}{\lambda^2 n^{2/3}}.
    \label{eq:transcond}
\end{equation}
In the limit that the first term on the left side is small compared to
the second (e.g. for decoupled relics),
the quantum liquid condition is:
\begin{equation}
    \sigma < \frac{\lambda(1-v^2)}{n^{2/3}}.
    \label{eq:relictranscond}
\end{equation}

Thus, for any decoupled cosmological relic, it becomes a quantum liquid
when its cross section is approximately less than the square of the
inter-particle separation.  This occurs faster for non-relativistic
relics $v \to 0$ than relativistic ones $v \to 1$, and can be delayed if
collisions are ``well-localized'' relative to the inter-particle
separation ($\lambda \to 0$).

This condition (Eq.\ref{eq:relictranscond}) is extremely well satisfied
for massive neutrinos and Weakly Interacting Massive Particle (WIMP)
dark matter, so that today, WIMPs and at least two neutrino mass
eigenstates are definitely quantum liquids.  

An important implication of this result is that non-relativistic relics
such as WIMP dark matter must be treated as quantum liquids.  The
phenomena currently attributed to dark matter can only be achieved by
a classical gas of particles which must satisfy $\Delta x(t) \ll
n(t)^{-1/3}$ 
One can see that under virtually any assumptions about Hubble expansion
and decoupling,
these
theories are only consistent if $M \gg M_{\rm Pl}$.  Such a heavy
object is very unlikely to be consistently described as a single quantum
particle.

If attractive contact interactions exist, the system will make a phase
transition to a super-fluid in exactly the same way as a BCS
superconductor or ${}^3$He.  For WIMP dark matter, the required contact
interaction occurs by integrating out any heavy particles which couple
to the WIMP to give a 4-point operator.  Collisions are so rare that
they can't break up the collective excitations of the super-fluid, and
the relevant condensation criterion is not given by the thermal
wavelength (Eq.~\ref{eq:nrtranstemp}) but rather the time-expanded wave
packet as in Eq.~\ref{eq:relictranscond}.  In the next section we show
that an attractive interaction always exists among fermions, though it
may be in a higher partial wave.

\section{The Kohn-Luttinger Effect}

Beyond wave-function overlap, a necessary condition for a super-fluid
state is the existence of a ground state with lower energy than the
original vacuum Lagrangian.  In the case of an attractive 4-fermion
interaction, there obviously exists a lower energy ground state where
the fermions bind into $s$-wave quasi-particles.  For WIMP dark matter
theories this is a possibility.

For the Standard Model (SM), neutrino self-interactions are
repulsive~\cite{caldi_cosmological_1999}.  However Kohn and Luttinger
showed that even a repulsive fermionic quantum liquid cannot behave as a
classical gas.  The reason is that at one loop, 4-point interactions
induce a singularity at the Fermi surface that is
attractive~\cite{PhysRevLett.15.524,KaganChubakov.47.525,efremov-2000-90}.
Since higher partial wave interactions are exponentially suppressed
relative to the $s$-wave, and this correction scales only as
$\ell^{-4}$, in terms of the partial wave number $\ell$.  For some large
$\ell$ this correction dominates.  For cosmological relics this occurs
already in the $p$ wave.

The relevant correction comes from an exchange (box) diagram and its
contribution to the BCS potential $V(x)$ in the $\ell$th partial wave is 
\begin{equation}
    \delta V_\ell = (-1)^{\ell+1} \frac{m p_F}{4 \pi^2}
    \frac{|V(\cos\theta=-1)|^2}{\ell^4}
    \label{eq:deltavl}
\end{equation}
where $p_F = (3 \pi n)^{1/3}$ and $V(\cos \theta)$ is the tree-level
potential evaluated on the Fermi surface.  This is attractive for odd
$\ell$,  The relevant infrared divergence occurs for $\cos \theta=-1$ and
corresponds to an exchange of the propagating neutrino with a background
neutrino.  The divergence occurs at $2 p_F$ because it occurs in the
internal loops, which contain two fermion propagators, both of which
must lie on the Fermi surface.

This potential is parametrically $\mathcal{O}(p_F^2 G_F^2)$.
Therefore this condensation is a much more important effect than
scattering, which is associated with the mean free path and is
$\mathcal{O}(p_F^5 G_F^2)$.  Note that $\delta V_1$ is also
parametrically the same order as Newton's constant $G_N$.

Therefore, an attractive self-interaction always exists in a neutrino or
fermionic WIMP fluid, regardless of the sign of the fundamental
interaction.  If the mass is sufficiently small so that the conditions
of the previous section are also satisfied, then such a cosmological relic
is a super-fluid today.  The two heavier neutrino species and WIMP dark
matter are super-fluids today.  Lighter species such the lightest
neutrino (if sufficiently light) would require an early-universe
analysis to determine if the conditions of the previous section can be
satisfied.

\section{Condensate Quantum Numbers}

A condensate will break Lorentz invariance, but if the
underlying theory is invariant, we can classify the condensates
by their Lorentz representation.
A Weyl fermion condenses as
$(\frac{1}{2},0) \otimes (\frac{1}{2},0) = (0,0) \oplus (1,0)$ according
to its representation under the spin Lorentz group.  
A $p$-wave condensate must contain a derivative, giving 
\begin{eqnarray}
    A_\mu(x,y) &=& \frac{i}{2}(\tilde\partial_\mu \chi\epsilon\xi - \chi \epsilon
    \tilde\partial_\mu \xi);\\
    E^a_\mu(x,y) &=& \frac{i}{2}(\tilde\partial_\mu \chi^\dagger\os^a\xi - \chi^\dagger
    \os^a \tilde\partial_\mu \xi),
    \label{eq:pwavecondensates}
\end{eqnarray}
where $\tilde \partial_\mu$ represents the deviation in momentum from
the Fermi surface, $p_0=0$, $|\vec p| = 2 p_F$, and we abbreviate $\chi
= \chi(x)$ and $\xi = \chi(y)$.  In condensed matter
nomenclature, these excitations are ``zero-sound''.

The four-point operator for these two condensates is the same
since they are related by a Fierz transformation, therefore we may write
it as
\begin{equation}
    \label{eq:effint}
    -\frac{g_Z^4 m p_F}{4 \pi^2 M_Z^4} \int_{xy}\left[(1-\eta_\nu)
    E^{a\dagger}_\mu E_a^\mu + \eta_\nu A_\mu^\dagger A^\mu\right],
\end{equation}
where 
\begin{equation}
    \eta_\nu = \frac{n_\nu - n_{\overline{\nu}}}{n_\nu +
    n_{\overline{\nu}}}
    \label{eq:etanu}
\end{equation}
is the asymmetry between neutrinos and anti-neutrinos.  After the phase
transition (Eq.\ref{eq:relictranscond}) has occurred, the original Fermi
gas is described by momentum distribution functions for $A_\mu$ and
$E^a_\mu$, rather than original one for free fermions.

The condensate $E^a_\mu$ contains both particles and antiparticles,
while $A_\mu$ contains only particles (or antiparticles).  Therefore,
$A_\mu$ only condenses among the unpaired particles that don't have an
antiparticle partner.  The Cosmic Neutrino Background (CNB) is expected
to contain very nearly equal numbers of neutrinos and anti-neutrinos.
The asymmetry $\eta_\nu$ is proportional to the baryon to photon ratio,
$\eta_b \sim 6 \times 10^{-10}$.
Therefore $E^a_\mu$ is the dominant condensate 
and the dynamics of $A_\mu$ are sub-leading so we will neglect them.
A right-handed neutrino state (if they are
Dirac) has interactions that are much weaker than the left-handed state,
and can be ignored.  Likewise, repulsive Majorana dark matter such as a
bino is usually not assumed to have any matter/antimatter asymmetry and
again can be treated as a single Weyl spinor super-fluid which condenses
into $E^a_\mu$.

\section{Lorentz Breaking}

The condensation of $A_\mu$ and $E^a_\mu$ breaks Poincar\'e invariance,
since both fields have Lorentz indices, and the neutrinos should have a
spatially varying density distribution.  This symmetry breaking is
dynamical and spontaneous, due to the condensation of a physical
background; the SM is Poincar\'e and Lorentz invariant.  As a
consequence of the symmetry breaking, both have corresponding Goldstone
bosons, which are long wavelength fluctuations about the expectation
values for $A_\mu$ and $E^a_\mu$.

Neutrino self-interactions are mediated by the $Z^0$ boson.  In the
Feynman gauge we may write the tree level effective 4-point operator as
\begin{equation}
    - \frac{g_Z^2}{2 M_Z^2} \int_{xy} \left\{ 
    \chi^\dagger \overline{\sigma}^a \chi
    \xi^\dagger \overline{\sigma}_a \xi \right\}.
    \label{eq:fourfermi}
\end{equation}
This interaction has the enhanced symmetry $SO(3,1)\times SO(3,1)$.  The
only term that breaks this enhanced symmetry is the fermion's kinetic
term, which ties together a derivative and a gamma or sigma matrix of
the spin Lorentz group:
\begin{equation}
    i \int_x \chi^\dagger \overline{\sigma}^\mu\partial_\mu \chi
    = \int_{xy} E^a_\mu \delta_a^\mu \delta^4 (x-y).
    \label{eq:kineticvev}
\end{equation}
However this term is a tadpole for the condensate $E^a_\mu$.  As such,
when $E^a_\mu$ condenses, the field must be shifted
$E^a_\mu \to \tilde E^a_\mu + \delta^a_\mu \delta^4(x-y)$ 
to remove this tadpole, and
$\tilde E^a_\mu$ is the order parameter of the $SO(3,1)\times SO(3,1)$
symmetry breaking.  In the limit that $\tilde E^a_\mu \to 0$, the
effective action has this enhanced symmetry (and the fermion has no
kinetic energy).

A free fermion $\psi(x)$ transforms with two Lorentz symmetries.  The
first is defined on the coordinates of space-time, with the generators 
\begin{equation}
L_{\mu\nu} = i(x_\mu \partial_\nu - x_\nu \partial_\mu).
\label{eq:Lmunu}
\end{equation}
Under this
symmetry $\psi$ transforms as a scalar.  The second Lorentz symmetry is
defined with the generators 
\begin{equation}
    S_{ab} = \frac{i}{2} (\gamma_a \gamma_b - \gamma_b \gamma_a),
    \label{eq:Sab}
\end{equation}
under which $\psi$ transforms in the $1/2$
(spinor) representation.
Normally we consider these to be two different representations of the
same $SO(3,1)$ Lorentz symmetry.  
The SM Lagrangian is not symmetric under both
groups separately.  We write Greek indices for the space-time
Lorentz group, and Roman indices for the spinor Lorentz group to
indicate the difference.  Since both groups contain the Minkowski metric
$\eta_{\mu\nu}$ and $\eta_{ab}$, we will use this to raise and lower
indices.  We can define the mixed generators
\begin{equation}
M_{\mu\nu} = L_{\mu\nu} + S_{ab}e^a_\mu e^b_\nu; \qquad
N_{\mu\nu} = L_{\mu\nu} - S_{ab}e^a_\mu e^b_\nu 
\label{mixedgenerators}
\end{equation}
where $e^a_\mu = \langle \tilde E^a_\mu \rangle \simeq \delta^a_\mu$.
The new operator $N_{\mu\nu}$ is the
broken generator, and corresponds for a massless fermion to local
violations of being in a helicity eigenstate.  
A plane wave could be a helicity eigenstate, but a localized
state is not an energy or momentum eigenstate, and therefore is also
cannot be a helicity eigenstate unless it is completely delocalized.  
Thus $e^a_\mu$ is the order parameter of the $SO(3,1)\times SO(3,1)\to
SO(3,1)$ symmetry breaking.

By Goldstone's theorem, a vacuum expectation value for $\tilde E^a_\mu$ not
only breaks this symmetry but also generates Goldstone bosons from the
broken symmetry generators.  Here care must be taken because the number
of Goldstones is not the same as the number of broken generators,
because the broken symmetry is a space-time
symmetry~\cite{Goldstone:1961eq,Goldstone:1962es,Low:2001bw}.

The Goldstones carry a representation of the unbroken
group $M_{\mu\nu}$.  The field $\tilde E^a_\mu$ however carries an index
of both the original groups.  The propagating Goldstone is
\begin{equation}
    g_{\mu\nu} = \tilde E^a_\mu \tilde E^b_\nu \eta_{ab}
    \label{eq:graviton}
\end{equation}
which we identify as spin-2 graviton under $M_{\mu\nu}$.  This should be
familiar from the Palatini formalism for quantizing gravity, if we
identify $\tilde E^a_\mu$ as the vierbein (tetrad).

The gravitational theory arising here does not conflict with the
Weinberg-Witten Theorem because of the presence of a physical
background, and consequently this emergent gravitational theory isn't
diffeomorphism invariant~\cite{Weinberg:1980kq}. There are many ways to
see this, but in particular, the Lorentz symmetry is not exact in the
gravitational theory, spatial variations of $p_F$ lead to a
spatially varying interaction strength (Eq.\ref{eq:effint}), and the
emergent vierbein (Eq.\ref{eq:pwavecondensates}) is nonlocal.

From here one can almost directly follow the program of ``Spinor
Gravity''~\cite{hebecker_spinor_2003,wetterich_gravity_2003}, with the
exception that due to the Lorentz symmetry breaking, we have the metric
$\eta_{\mu\nu}$ with which to tie up spacetime indices, which gives rise
to a spin connection which was absent in ``Spinor Gravity''.  The
existence of $\eta_{\mu\nu}$ implies more invariants as well.  

\section{Conclusions} 
We have shown that massive cosmological relics are not classical gasses.
If they have attractive interactions or are fermions, they instead are a
super-fluid.  This implies that WIMP dark matter scenarios are
inconsistent: WIMPs cannot both be decoupled and localized for the age
of the universe.

Cosmic background neutrinos must exist. They are a super-fluid, and their
self-interactions are a gravitational theory.  These dynamics
arise in the SM, which is a renormalizable quantum field theory.  We
suggest that this may actually be the gravity that we observe.

\section{Acknowledgements}
We thank 
Bruce Campbell, 
Steve Carlip,
Jessica De Haene, 
Francois Gelis, 
Patrick Huber, 
Nemanja Kaloper, 
Alessio Notari,
Thomas Schwetz, 
Steve Sekula, 
Aleksi Vuorinen,
Edward Witten, and 
Jure Zupan for useful comments.


\bibliographystyle{unsrt}
\bibliography{ewgravity}

\end{document}